\documentclass[aps,prd,onecolumn, groupedaddress, showpacs, nofootinbib, amssymb, preprint]{revtex4-2}
\usepackage[dvips]{graphicx}
\usepackage{amssymb}
\usepackage{amsmath}
\usepackage{graphicx,,color}
\usepackage{amsfonts}
\usepackage{bm}
\usepackage{cancel}
\usepackage{comment}
\usepackage{floatflt}
\usepackage{slashed}

\newcommand\be{\begin{equation}}
\newcommand\ee{\end{equation}}

\newcommand\e{\mathrm{e}}

\allowdisplaybreaks[4]
\begin{document}

\preprint{KEK-TH-2661, KEK-Cosmo-0361}
\title{
Black holes, photon sphere, and cosmology in ghost-free $f(\mathcal{G})$ gravity
}

\author{Shin'ichi~Nojiri$^{1,2}$}\email{nojiri@gravity.phys.nagoya-u.ac.jp}
\author{S.~D.~Odintsov$^{3,4}$}\email{odintsov@ice.csic.es}

\affiliation{ $^{1)}$ Theory Center, High Energy Accelerator Research Organization (KEK), Oho 1-1, Tsukuba, Ibaraki 305-0801, Japan \\
$^{2)}$ Kobayashi-Maskawa Institute for the Origin of Particles and the Universe, Nagoya University, Nagoya 464-8602, Japan \\
$^{3)}$ Institute of Space Sciences (ICE, CSIC) C. Can Magrans s/n, 08193 Barcelona, Spain \\
$^{4)}$ ICREA, Passeig Lluis Companys, 23, 08010 Barcelona, Spain
}

\begin{abstract}
The improved version of ghost-free $f(\mathcal{G})$ gravity introduced in Phys.~Lett.~B \textbf{631} (2005), 1-6 is proposed and investigated. 
It is demonstrated that improved ghost-free $f(\mathcal{G})$ gravity may be consistently applied to describe the expanding universe (with horizon) 
as well as the static spacetime like black holes. 
Interestingly, such a theory looks like a close avatar of scalar-Einstein-Gauss-Bonnet gravity with the only difference that the scalar field is not a dynamical one 
so that many predictions of ghost-free $f(\mathcal{G})$ gravity are very similar to that of sEGB gravity.

We discuss the gravitational wave in the improved model with the condition that the propagating speed of the gravitational wave 
is identical to that of the propagation speed of light. 
A model that describes inflation in the early universe and could satisfy the observational constraints is also constructed. 
The solution of ghost-free $f(\mathcal{G})$ gravity realising the given static spacetime with spherical symmetry is proposed. 
Some of such solutions realise explicitly the Reissner-Nordstr\"{o}m black hole or the Hayward black hole, which is a regular black hole without curvature singularity, 
We also construct a black hole in our theory which contains the Arnowitt-Deser-Misner (ADM) mass, the horizon radius, and the radius of the photon sphere 
as independent parameters. 
The radius of the black hole shadow in this model as well as photon sphere radius are estimated. 
It is shown that there exists a parameter region which satisfies the constraints coming from Event Horizon Telescope observations. 
Furthermore, we construct model where the radius of the black hole shadow or photon orbit is smaller than  the horizon radius 
supposed from the ADM mass. 

\end{abstract}

\maketitle

\section{Introduction\label{SecI}}

Different models of modified gravity (for review, see \cite{Capozziello:2011et, Nojiri:2010wj, Nojiri:2017ncd, Faraoni:2010pgm} 
have been proposed to describe consistently the early-time and late-time acceleration of the universe. 
One well-studied class of the modified gravities is the scalar-Einstein-Gauss-Bonnet (sEGB) gravity theory~\cite{Nojiri:2005vv, Nojiri:2006je}, 
which includes a propagating scalar field $\phi$ with potential and the Gauss-Bonnet topological invariant coupled with the scalar field in the action. 
This model can be interpreted as a generalisation of the model given by the $\alpha'$ corrected effective action in superstring theories~\cite{Gross:1986mw}. 
This model includes a propagating scalar mode but by removing the kinetic term of the scalar, the scalar field can be regarded as an auxiliary field. 
By integrating the auxiliary scalar field, there appears a function $f(\mathcal{G})$ of the Gauss-Bonnet invariant $\mathcal{G}$ in the action. 
Such a theory is called $f(\mathcal{G})$ gravity. 
It has been first introduced in~\cite{Nojiri:2005jg} and futher studied in Refs.~\cite{Cognola:2006eg, Leith:2007bu, Li:2007jm, Kofinas:2014owa, Zhou:2009cy}. 
However, it has been found that $f(\mathcal{G})$ gravity includes ghosts. 
Classically, the kinetic energy of the ghosts is unbounded below and as a quantum theory, the model includes negative norm states. 
The ghosts in quantum theory conflict with the so-called Copenhagen interpretation in quantum theory. 

The problem of the ghost has been solved in \cite{Nojiri:2018ouv} where a ghost-free version of the theory was formulated. 
This was done by using of the Lagrange multiplier constraint (for the review of theories with such constraint, see \cite{Sebastiani:2016ras, Lim:2010yk}). 
The corresponding theory looks like a ghost-free avatar of sEGB gravity. 
Indeed, the scalar degree of freedom in the scalar-Einstein-Gauss-Bonnet gravity is eliminated by imposing a constraint so 
that scalars become non-dynamical in ghost-free $f(\mathcal{G})$ gravity. 
Note that this constraint makes the situation complicated, that is, we need to impose different kinds of constraints in the region 
where the derivative of the scalar field $\partial_\mu\phi$ is time-like and the region where it is space-like~\cite{Nojiri:2022cah}. 
Therefore the model which works well in the expanding universe seems to be not fully consistent in the static spacetime especially when there is a horizon in the spacetime. 
This problem can be solved by introducing a function of the scalar field in the constraint. 
This improved constraint gives original ones in the corresponding gravity if we only consider one of the regions where $\partial_\mu\phi$ is space-like or time-like but 
the modified constraint is always valid in both of the two regions. 
Such a formulation was recently applied to cosmology and black hole physics, especially the black hole shadow in \cite{Nojiri:2024txy}. 

In this paper, we apply the improved formulation for the constraint in the ghost-free $f(\mathcal{G})$ gravity model. 
As a result, we can discuss the cosmology of the expanding universe and static black holes naturally and consistently. 
Moreover, we can see explicitly the difference between ghost-free $f(\mathcal{G})$ and sEGB gravity which comes from differences 
in the scalars (non-dynamical ones and dynamical ones). 
The paper is organised as follows.
In the next section, we review the ghost-free $f(\mathcal{G})$ gravity and show how the constraint is improved. 
In Section~\ref{SecIII}, we consider the gravitational wave in this model and show the condition that the propagating speed of the gravitational wave is equal with 
that of the propagation speed of light is just the same as in sEGB gravity.
In Section~\ref{SecIV}, we discuss cosmology and we propose a model which describes the inflation in the early universe. 
In Section~\ref{SecV}, we propose the solution for several kinds of black hole spacetimes, including the Reissner-Nordstr\"{o}m black hole 
and the Hayward black hole, which is known to be a regular black hole. 
The Reissner-Nordstr\"{o}m black hole in this section has a hair of the scalar field instead of the electromagnetic field. 
In Sectin~\ref{SecVI}, we investigate black hole shadow in a model which includes the Arnowitt-Deser-Misner (ADM) mass, the horizon radius, 
and the radius of the photon sphere as independent parameters. 
We find the radius of the black hole shadow in this model and consider the parameter regions consistent with the observations. 
We also construct models where the radius of the black hole shadow or photon orbit is smaller than the radius appearing as the horizon radius 
supposed from the ADM mass. 
The last section is devoted to the summary and discussions. 

\section{New ghost-free Einstein-$f(\mathcal{G})$ gravity\label{SecII}}

The sEGB gravity theory~\cite{Nojiri:2005vv, Nojiri:2006je} is described by the following action, 
\begin{align}
\label{I8one}
S_{\mathrm{GB}\phi} = \int d^4 x \sqrt{-g} & \left\{ \frac{R}{2\kappa^2}
 - \frac{1}{2} \partial_\mu \phi \partial^\mu \phi - V (\phi) - \xi(\phi) \mathcal{G} + \mathcal{L}_\mathrm{matter} \right\}\, .
\end{align}
Here $V(\phi)$ is the potential of a scalar field $\phi$, $\xi(\phi)$ is also a function of $\phi$, and 
we denote the Lagrangian density of the matter by $\mathcal{L}_\mathrm{matter}$.
The Gauss-Bonnet topological invariant $\mathcal{G}$ is defined by,
\begin{align}
\label{eq:GB}
\mathcal{G} = R^2-4R_{\alpha \beta}R^{\alpha \beta}+R_{\alpha \beta \rho \sigma}R^{\alpha \beta \rho \sigma}\, .
\end{align}
The model (\ref{I8one}) is motivated by the $\alpha'$ corrected effective action in superstring theories~\cite{Gross:1986mw}. 

We may delete the kinetic term of $\phi$, $ - \frac{1}{2} \partial_\mu \phi \partial^\mu \phi$, then $\phi$ becomes an auxiliary field. 
The variation of the action (\ref{I8one}) without the kinetic term gives the equation, $0=V' (\phi) + \xi'(\phi) \mathcal{G}$, which can be 
algebraically solved with respect to $\phi$ as a function of the Gauss-Bonnet invariant, $\phi=\phi(G)$, in principle. 
We may substitute the obtained expression into the action (\ref{I8one}) which does not have the kinetic term. 
Then the resulting action describe the $f(\mathcal{G})$ gravity 
model~\cite{Nojiri:2005jg, Cognola:2006eg, Leith:2007bu, Li:2007jm, Kofinas:2014owa, Zhou:2009cy}, 
\begin{equation}
\label{GB1b} 
S=\int d^4x\sqrt{-g} \left(\frac{1}{2\kappa^2}R + f(\mathcal{G}) + \mathcal{L}_\mathrm{matter}\right)\, , \quad 
f(\mathcal{G})= - V \left( \phi\left(\mathcal{G}\right) \right) - \xi \left( \phi\left(\mathcal{G}\right) \right) \mathcal{G} \, .
\end{equation}
It is known, however, that ghosts appear in the $f(\mathcal{G})$ gravity \cite{DeFelice:2009ak}. 
The kinetic energy of ghosts is unbounded below in the classical theory and in the quantum theory, the ghosts generate negative norm states, 
which conflicts with the Copenhagen interpretation. 
Nevertheless, the Fadeev-Popov ghosts in the gauge theories could be well-known and tractable ghosts~\cite{Kugo:1977zq, Kugo:1979gm}.

In order to solve the problem of the ghost, a new type of the Einstein--Gauss-Bonnet modified gravity model 
has been proposed in \cite{Nojiri:2018ouv}. 
The action of the model is given by 
\begin{align}
\label{gfEGB1}
S_{\mathrm{GB}\phi} = \int d^4 x \sqrt{-g} & \left\{ \frac{R}{2\kappa^2} + \lambda \left( \partial_\mu \phi \partial^\mu \phi - \mu^4 \right) 
 - \frac{1}{2} \partial_\mu \phi \partial^\mu \phi - V (\phi) - \xi(\phi) \mathcal{G} + \mathcal{L}_\mathrm{matter} \right\}\, .
\end{align}
Here $\mu$ is a parameter with a mass dimension and $\lambda$ is the Lagrange multiplier field. 
By the variation of the action (\ref{gfEGB1}) with respect to $\lambda$, we obtain the following constraint, 
\begin{align}
\label{gfEGB2}
0 = \partial_\mu \phi \partial^\mu \phi - \mu^4 \, .
\end{align}
This constraint makes the scalar field non-dynamical and any ghost is not included in the model (\ref{gfEGB1})~\cite{Nojiri:2018ouv}. 

This constraint, however, makes the situation complicated when we consider the spacetime with the horizon as in the black hole. 
For the demonstration, we consider the static and spherically symmetric spacetime, 
\begin{align}
\label{metD}
ds^2= - \e^{2\nu(r)} dt^2 + \e^{2\eta(r)} dr^2 + r^2 d{\Omega_2}^2\, .
\end{align}
Here $d{\Omega_2}^2$ is the line element of the two-dimensional unit sphere. 
If we may also assume the scalar field $\phi$ only depends on the radial coordinate $r$, $\phi=\phi(r)$ because the spacetime is static, 
the constraint (\ref{gfEGB2}) has the following form
\begin{align}
\label{cnstrnt1}
\e^{-2\eta(r)} \left( \phi' \right)^2 = \mu^4 \, .
\end{align}
Note there is no solution in the equation~(\ref{cnstrnt1}) if $\e^{-2\eta(r)}$ is negative, $\e^{-2\eta(r)}<0$.
In the case of black hole geometry, $\e^{2\nu(r)}$ vanishes and changes its signature at the horizon.
It should be noted that $\e^{2\eta(r)}$ also must vanish at the horizon to avoid the curvature singularity. 
Therefore due to the constraint (\ref{gfEGB2}), we cannot realise the black hole geometry with the horizon(s) 
if the solution is static and $\phi$ only depends on $r$. 

Furthermore, we may consider the spatially flat Friedmann-Lema\^{i}tre-Robertson-Walker (FLRW) Universe, 
\begin{align}
\label{FLRW}
ds^2= -dt^2 + a(t)^2\sum_{i=1,2,3} \left(dx^i\right)^2 \, .
\end{align}
Here $t$ is the cosmological time, and $a(t)$ denotes the scale factor and we often use the Hubble rate $H= \frac{\dot a}{a}$. 
In the homogeneous and isotropic universe, a natural assumption could be that $\phi$ only depends on the cosmological time $t$. 
Under the assumption, the constraint (\ref{gfEGB2}) is given by, 
\begin{align}
\label{FLRWcons}
0 = {\dot \phi}^2 + \mu^4 \, ,
\end{align}
where there is no real solution for $\dot\phi$. 
Instead of (\ref{gfEGB2}), in the FLRW universe, the constraint should be changed as, 
\begin{align}
\label{gfEGB2FLRW}
0 = \partial_\mu \phi \partial^\mu \phi - \mu^4 \, ,
\end{align}
which makes the assumption $\phi=\phi(t)$ consistent. 
On the other hand, the constraint (\ref{gfEGB2FLRW}) cannot be satisfied in the static spacetime, where $\phi$ does not depend on time coordinate. 

This problem has been solved by modifying the constraint (\ref{gfEGB2}) by introducing a function $\omega(\phi)$ 
as follows~\cite{Nojiri:2022cah}, 
\begin{align}
\label{gfEGB3}
0 = \omega(\phi) \partial_\mu \phi \partial^\mu \phi - \mu^4 \, .
\end{align}
When $\omega(\phi)$ is positive, by redefining a scalar field $\tilde\phi$ as $\tilde\phi = \int d\phi \sqrt{\omega(\phi)}$, 
we find that the constraint (\ref{gfEGB3}) is reduced to the form of (\ref{cnstrnt1}),
\begin{align}
\label{lambdavar2}
g^{\rho \sigma}\partial_\rho \tilde\phi \partial_\sigma \tilde\phi= \mu^4 \, .
\end{align}
The signature of $\omega(\phi)$ can be, however, changed in general.
When the spacetime is given by (\ref{metD}), by assuming $\phi=\phi(r)$, 
instead of (\ref{cnstrnt1}), we find the constraint (\ref{gfEGB3}) has the following form,
\begin{align}
\label{cnstrnt2}
\e^{-2\eta(r)} \omega(\phi) \left( \phi' \right)^2 = \mu^4 \, .
\end{align}
Therefore, if we have a solution of $\phi$ where $\omega(\phi)$ is positive when $\e^{-2\eta(r)}$ is positive 
and $\omega(\phi)$ is negative when $\e^{-2\eta(r)}$ is negative,
we find that the constraint (\ref{cnstrnt2}) is consistent even inside the horizon. 
When $\omega(\phi)$ is negative, by redefining a scalar field $\hat\phi$ by $\hat\phi=\int d\phi \sqrt{-\omega(\phi)}$, instead of (\ref{lambdavar2}), 
the following constraint is obtained, 
\begin{align}
\label{lambdavar2BB}
g^{\rho \sigma}\partial_\rho \hat\phi \partial_\sigma \hat\phi= - \mu^4 \, .
\end{align}
Therefore we find that by introducing the function $\omega(\phi)$, in the identical model, it becomes possible to consider both the spherically symmetric 
and static spacetime and the expanding universe like the FLRW universe in (\ref{FLRW}), where $\partial_\mu \phi$ could be time-like. 

A simple choice is given by 
\begin{align}
\label{ex1}
\omega(\phi) = \frac{1}{\phi} \, .
\end{align}
Near the horizon, $\e^{-2\eta(r)}$ in (\ref{metD}) should behave as
\begin{align}
\label{ex2}
\e^{-2\eta(r)}(r) \sim b_0 \left( r - r_\mathrm{h} \right) \, .
\end{align}
Here $r_\mathrm{h}$ is the radius of the horizon and $b_0$ is a positive constant.
Then a solution of (\ref{cnstrnt2}) with (\ref{ex1}) is given by 
\begin{align}
\label{ex3}
\phi \sim \frac{\mu^4 \left( r - r_\mathrm{h}\right)}{b_0} \, .
\end{align}
Then the scalar $\phi$ and therefore $\omega(\phi)$ change the sign at the horizon and 
we find Eq.~(\ref{cnstrnt2}) is consistent even inside the horizon.

In the case that there are several horizons, we may need to extend the choice in (\ref{ex1}). 
A simple extension is given by, 
\begin{align}
\label{ex4}
\omega(\phi) = \mu^4 \e^{2\eta(r=\phi)} \, .
\end{align}
For the choice, the solution of (\ref{cnstrnt2}) is simply given by 
\begin{align}
\label{ex5}
\phi = r \, .
\end{align}
Therefore it is clear that the problem is solved by the choice of (\ref{ex4}). 
One might think, however, that the choice (\ref{ex4}) could be rather artificial because it looks like we have assumed the solution from the beginning. 
Anyway, the possibility of the choice (\ref{ex4}) shows that a model gives the solution of Eq.~(\ref{cnstrnt2}). 

In the following, we consider the following model, 
\begin{align}
\label{gfEGB4}
S_{\mathrm{GB}\phi} = \int d^4 x \sqrt{-g} &\, \left\{ \frac{R}{2\kappa^2} + \lambda \left( \omega(\phi) \partial_\mu \phi \partial^\mu \phi - \mu^4 \right) 
\right. \nonumber \\
&\, \left. - \frac{1}{2} \partial_\mu \phi \partial^\mu \phi - V (\phi) - \xi(\phi) \mathcal{G} + \mathcal{L}_\mathrm{matter} \right\}\, , 
\end{align}
which could be regarded as the new ghost-free Einstein-$f(\mathcal{G})$ gravity. 

The variation of the action (\ref{gfEGB4}) with respect to the metric $g_{\mu\nu}$ gives the following equation corresponding to the Einstein equation, 
\begin{align}
\label{gb4bD4one}
0= &\, \frac{1}{2\kappa^2}\left(- R_{\mu\nu} + \frac{1}{2} g_{\mu\nu} R\right)
+ \frac{1}{2} g_{\mu\nu} \left\{ \lambda \left( \omega(\phi) \partial_\mu \phi \partial^\mu \phi - \mu^4 \right)
 - \frac{1}{2} \partial_\rho \phi \partial^\rho \phi - V (\phi)\right\} \nonumber \\
&\, - \lambda \omega(\phi) \partial_\mu \phi \partial_\nu \phi 
+ \frac{1}{2} \partial_\mu \phi \partial_\nu \phi \nonumber \\
&\, - 2 \left( \nabla_\mu \nabla_\nu \xi(\phi)\right)R
+ 2 g_{\mu\nu} \left( \nabla^2 \xi(\phi)\right)R
+ 4 \left( \nabla_\rho \nabla_\mu \xi(\phi)\right)R_\nu^{\ \rho}
+ 4 \left( \nabla_\rho \nabla_\nu \xi(\phi)\right)R_\mu^{\ \rho} \nonumber \\
&\, - 4 \left( \nabla^2 \xi(\phi) \right)R_{\mu\nu}
 - 4g_{\mu\nu} \left( \nabla_\rho \nabla_\sigma \xi(\phi) \right) R^{\rho\sigma}
+ 4 \left(\nabla^\rho \nabla^\sigma \xi(\phi) \right) R_{\mu\rho\nu\sigma}
+ \frac{1}{2} T_{\mathrm{matter}\, \mu\nu} \, .
\end{align}
By varying the action with respect to $\phi$, we obtain the following field equation, 
\begin{align}
\label{I10one}
0 =&\, \lambda \omega'(\phi) \partial_\mu \phi \partial^\mu \phi - 2 \nabla^\mu \left( \lambda \omega \left(\phi \right) \partial_\mu \phi \right)
+ \nabla^\mu \partial_\mu \phi - V' - \xi' \mathcal{G} \, .
\end{align}
In (\ref{gb4bD4one}), $T_{\mathrm{matter}\, \mu\nu}$ is the energy-momentum tensor of the perfect matter fluid, 
which obeys the conservation law $0=\nabla^\mu T_{\mathrm{matter}\, \mu\nu}$. 
The variation of the action (\ref{gfEGB4}) with respect to $\lambda$ gives the constraint (\ref{cnstrnt2}). 
Because Eq.~(\ref{I10one}) can be obtained by using (\ref{gb4bD4one}) and the conservation law, 
we dont use Eq.~(\ref{I10one}) in the remaining part in this paper. 

\section{Gravitational Wave\label{SecIII}}

In this section, for the model in (\ref{gfEGB4}), we study the gravitational wave and find 
the condition that the propagating speed of the gravitational waves is identical to that of light in vacuum. 

In the case of the scalar--Einstein--Gauss-Bonnet theories, by the GW170817 neutron star merger event
\cite{TheLIGOScientific:2017qsa, Monitor:2017mdv, GBM:2017lvd}, we obtain a strong constraint 
on the coupling of the scalar field with the Gauss-Bonnet invariant, which is a function often denoted by $\xi(\phi)$ in (\ref{I8one}). 
After that, several scenarios of an Einstein-Gauss-Bonnet theory compatible with the GW170817 observation have been 
proposed~\cite{Odintsov:2020xji, Odintsov:2020sqy, Oikonomou:2021kql, Oikonomou:2022ksx}. 
Although the scenario works for the FLRW background spacetime in (\ref{FLRW}), 
any constraint cannot be satisfied in the static and spherically symmetric spacetime~\cite{Nojiri:2023jtf}. 
In this section, for the model in (\ref{gfEGB4}), we consider 
the condition that the propagating speed of the gravitational waves is equal to that of light in a vacuum. 

For the general variation of the metric $g_{\mu\nu}$ denoted by $h_{\mu\nu}$,
\begin{align}
\label{variation1}
g_{\mu\nu}\to g_{\mu\nu} + h_{\mu\nu}\, ,
\end{align}
we have the following equation in leading order in terms of $h_{\mu\nu}$, that is, $\mathcal{O}\left(h\right)$, 
\begin{align}
\label{variation2}
\delta\Gamma^\kappa_{\mu\nu} =&\, \frac{1}{2}g^{\kappa\lambda}\left(
\nabla_\mu h_{\nu\lambda} + \nabla_\nu h_{\mu\lambda} - \nabla_\lambda h_{\mu\nu}
\right)\, ,\nonumber \\
\delta R^\mu_{\ \nu\lambda\sigma}=&\, \nabla_\lambda \delta\Gamma^\mu_{\sigma\nu}
 - \nabla_\sigma \delta \Gamma^\mu_{\lambda\nu}\, ,\nonumber \\
\delta R_{\mu\nu\lambda\sigma}=&\, \frac{1}{2}\left[\nabla_\lambda \nabla_\nu h_{\sigma\mu}
 - \nabla_\lambda \nabla_\mu h_{\sigma\nu}
 - \nabla_\sigma \nabla_\nu h_{\lambda\mu}
 + \nabla_\sigma \nabla_\mu h_{\lambda\nu}
+ h_{\mu\rho} R^\rho_{\ \nu\lambda\sigma}
 - h_{\nu\rho} R^\rho_{\ \mu\lambda\sigma} \right] \, ,\nonumber \\
\delta R_{\mu\nu} =& \frac{1}{2}\left[\nabla^\rho\left(\nabla_\mu h_{\nu\rho}
+ \nabla_\nu h_{\mu\rho}\right) - \nabla^2 h_{\mu\nu}
 - \nabla_\mu \nabla_\nu \left(g^{\rho\lambda}h_{\rho\lambda}\right)\right] \nonumber \\
=&\, \frac{1}{2}\left[\nabla_\mu\nabla^\rho h_{\nu\rho}
+ \nabla_\nu \nabla^\rho h_{\mu\rho} - \nabla^2 h_{\mu\nu}
 - \nabla_\mu \nabla_\nu \left(g^{\rho\lambda}h_{\rho\lambda}\right)
 - 2R^{\lambda\ \rho}_{\ \nu\ \mu}h_{\lambda\rho}
+ R^\rho_{\ \mu}h_{\rho\nu} + R^\rho_{\ \mu}h_{\rho\nu} \right]\, ,\nonumber \\
\delta R =&\, -h_{\mu\nu} R^{\mu\nu} + \nabla^\mu \nabla^\nu h_{\mu\nu}
 - \nabla^2 \left(g^{\mu\nu}h_{\mu\nu}\right)\, , 
\end{align}
which gives the variation of the equation (\ref{gb4bD4one}) corresponding to the Einstein equation as follows, 
\begin{align}
\label{gb4bD4B}
0=&\, \left[ \frac{1}{4\kappa^2} R + \frac{1}{2} \left\{
 - \frac{1}{2} \partial_\rho \phi \partial^\rho \phi - V \right\}
 - 4 \left( \nabla_{\rho} \nabla_\sigma \xi \right) R^{\rho\sigma} \right] h_{\mu\nu} 
+ \bigg[ \left( \frac{1}{2} \lambda \omega(\phi) - \frac{1}{4} \right) g_{\mu\nu} \partial^\tau \phi \partial^\eta \phi
\nonumber \\
&\, - 2 g_{\mu\nu} \left( \nabla^\tau \nabla^\eta \xi\right)R
 - 4 \left( \nabla^\tau \nabla_\mu \xi\right)R_\nu^{\ \eta} - 4 \left( \nabla^\tau \nabla_\nu \xi\right)R_\mu^{\ \eta}
+ 4 \left( \nabla^\tau \nabla^\eta \xi \right)R_{\mu\nu} \nonumber \\
&\, + 4g_{\mu\nu} \left( \nabla^\tau \nabla_\sigma \xi \right) R^{\eta\sigma}
+ 4g_{\mu\nu} \left( \nabla_\rho \nabla^\tau \xi \right) R^{\rho\eta}
 - 4 \left(\nabla^\tau \nabla^\sigma \xi \right) R_{\mu\ \, \nu\sigma}^{\ \, \eta}
 - 4 \left(\nabla^\rho \nabla^\tau \xi \right) R_{\mu\rho\nu}^{\ \ \ \ \eta}
\bigg] h_{\tau\eta} \nonumber \\
&\, + \frac{1}{2}\left\{ 2 \delta_\mu^{\ \eta} \delta_\nu^{\ \zeta} \left( \nabla_\kappa \xi \right)R
 - 2 g_{\mu\nu} g^{\eta\zeta} \left( \nabla_\kappa \xi \right)R
 - 4 \delta_\rho^{\ \eta} \delta_\mu^{\ \zeta} \left( \nabla_\kappa \xi \right)R_\nu^{\ \rho}
 - 4 \delta_\rho^{\ \eta} \delta_\nu^{\ \zeta} \left( \nabla_\kappa \xi \right)R_\mu^{\ \rho} \right. \nonumber \\
&\, \left. + 4 g^{\eta\zeta} \left( \nabla_\kappa \xi \right) R_{\mu\nu}
+ 4g_{\mu\nu} \delta_\rho^{\ \eta} \delta_\sigma^{\ \zeta} \left( \nabla_\kappa \xi \right) R^{\rho\sigma}
 - 4 g^{\rho\eta} g^{\sigma\zeta} \left( \nabla_\kappa \xi \right) R_{\mu\rho\nu\sigma}
\right\} g^{\kappa\lambda}\left( \nabla_\eta h_{\zeta\lambda} + \nabla_\zeta h_{\eta\lambda} - \nabla_\lambda h_{\eta\zeta} \right) \nonumber \\
&\, + \left\{ \frac{1}{4\kappa^2} g_{\mu\nu} - 2 \left( \nabla_\mu \nabla_\nu \xi\right) + 2 g_{\mu\nu} \left( \nabla^2\xi\right) \right\}
\left\{ -h_{\mu\nu} R^{\mu\nu} + \nabla^\mu \nabla^\nu h_{\mu\nu} - \nabla^2 \left(g^{\mu\nu}h_{\mu\nu}\right) \right\} \nonumber \\
&\, + \frac{1}{2}\left\{ \left( - \frac{1}{2\kappa^2} - 4 \nabla^2 \xi \right) \delta^\tau_{\ \mu} \delta^\eta_{\ \nu}
+ 4 \left( \nabla_\rho \nabla_\mu \xi\right) \delta^\eta_{\ \nu} g^{\rho\tau}
+ 4 \left( \nabla_\rho \nabla_\nu \xi\right) \delta^\tau_{\ \mu} g^{\rho\eta}
 - 4g_{\mu\nu} \nabla^\tau \nabla^\eta \xi \right\} \nonumber \\
&\, \qquad \times \left\{\nabla_\tau\nabla^\phi h_{\eta\phi}
+ \nabla_\eta \nabla^\phi h_{\tau\phi} - \nabla^2 h_{\tau\eta}
 - \nabla_\tau \nabla_\eta \left(g^{\phi\lambda}h_{\phi\lambda}\right)
 - 2R^{\lambda\ \phi}_{\ \eta\ \tau}h_{\lambda\phi}
+ R^\phi_{\ \tau}h_{\phi\eta} + R^\phi_{\ \tau}h_{\phi\eta} \right\} \nonumber \\
&\, + 2 \left(\nabla^\rho \nabla^\sigma \xi \right)
\left\{ \nabla_\nu \nabla_\rho h_{\sigma\mu}
 - \nabla_\nu \nabla_\mu h_{\sigma\rho}
 - \nabla_\sigma \nabla_\rho h_{\nu\mu}
 + \nabla_\sigma \nabla_\mu h_{\nu\rho}
+ h_{\mu\phi} R^\phi_{\ \rho\nu\sigma}
 - h_{\rho\phi} R^\phi_{\ \mu\nu\sigma} \right\} \nonumber \\
&\, + \frac{1}{2}\frac{\partial T_{\mathrm{matter}\,
\mu\nu}}{\partial g_{\tau\eta}}h_{\tau\eta} \, .
\end{align}
Here we assumed that the perfect matter fluids are minimally coupled with gravity and we used (\ref{gfEGB3}). 
We now choose the following gauge-fixing condition, 
\begin{align}
\label{gfc}
0=\nabla^\mu h_{\mu\nu}\, .
\end{align}
Because we are interested in the standard gravitational wave, that is, the massless spin-two tensor mode, 
we may assume the traceless condition,
\begin{align}
\label{ce}
0=g^{\mu\nu} h_{\mu\nu} \, .
\end{align}
In this paper, we do not consider the perturbation of the scalar mode in the metric like the trace part, which may couple with the scalar field $\phi$ 
in general (see \cite{Dalang:2020eaj} for example) because we are interested in the usual gravitational wave. 
We should note that it is difficult to detect the scalar wave in the present gravitational wave detector like LIGO. 
For the usual scalar field as in the scalar--Einstein--Gauss-Bonnet gravity, as long as we consider the leading order of the perturbation, 
the massless spin-two mode does not mix with the scalar mode,
which is a massive spin-zero mode although the second-order perturbation of the scalar field plays the role of the source of the gravitational wave.
The traceless condition (\ref{ce}) makes the massless spin-two mode decouple with the massive spin-zero mode in the leading order. 
Furthermore in the case of the ghost-free $f(\mathcal{G})$ gravity model in this paper, due to the constraint (\ref{gfEGB3}), the scalar field $\phi$ 
does not propagate. 
Therefore even if there is a mixing between the scalar mode of the metric and the scalar field $\phi$, there could not appear scalar wave unlike in the case of sEGB gravity. 

Then Eq.~(\ref{gb4bD4B}) is reduced into the following form, 
\begin{align}
\label{gb4bD4BoneGW}
0=&\, \left[ \frac{1}{4\kappa^2} R + \frac{1}{2} \left\{
 - \frac{1}{2} \partial_\rho \phi \partial^\rho \phi - V \right\}
 - 4 \left( \nabla_\rho \nabla_\sigma \xi \right) R^{\rho\sigma} \right] h_{\mu\nu} 
+ \bigg[ \left( \frac{1}{2} \lambda \omega(\phi) - \frac{1}{4} \right) g_{\mu\nu} \partial^\tau \phi \partial^\eta \phi 
\nonumber \\
&\, - 2 g_{\mu\nu} \left( \nabla^\tau \nabla^\eta \xi\right)R
 - 4 \left( \nabla^\tau \nabla_\mu \xi\right)R_\nu^{\ \eta} - 4 \left( \nabla^\tau \nabla_\nu \xi\right)R_\mu^{\ \eta}
+ 4 \left( \nabla^\tau \nabla^\eta \xi \right)R_{\mu\nu} \nonumber \\
&\, + 4g_{\mu\nu} \left( \nabla^\tau \nabla_\sigma \xi \right) R^{\eta\sigma}
+ 4g_{\mu\nu} \left( \nabla_{\rho} \nabla^\tau \xi \right) R^{\rho\eta}
 - 4 \left(\nabla^\tau \nabla^\sigma \xi \right) R_{\mu\ \, \nu\sigma}^{\ \, \eta}
 - 4 \left(\nabla^\rho \nabla^\tau \xi \right) R_{\mu\rho\nu}^{\ \ \ \ \eta}
\bigg\} h_{\tau\eta} \nonumber \\
&\, + \frac{1}{2}\left\{ 2 \delta_\mu^{\ \eta} \delta_\nu^{\ \zeta} \left( \nabla_\kappa \xi \right)R
 - 4 \delta_\rho^{\ \eta} \delta_\mu^{\ \zeta} \left( \nabla_\kappa \xi \right)R_\nu^{\ \rho}
 - 4 \delta_\rho^{\ \eta} \delta_\nu^{\ \zeta} \left( \nabla_\kappa \xi \right)R_\mu^{\ \rho} \right. \nonumber \\
&\, \left. + 4g_{\mu\nu} \delta_\rho^{\ \eta} \delta_\sigma^{\ \zeta} \left( \nabla_\kappa \xi \right) R^{\rho\sigma}
 - 4 g^{\rho\eta} g^{\sigma\zeta} \left( \nabla_\kappa \xi \right) R_{\mu\rho\nu\sigma}
\right\} g^{\kappa\lambda}\left( \nabla_\eta h_{\zeta\lambda} + \nabla_\zeta h_{\eta\lambda} - \nabla_\lambda h_{\eta\zeta} \right) \nonumber \\
&\, - \left\{ \frac{1}{4\kappa^2} g_{\mu\nu} - 2 \left( \nabla_\mu \nabla_\nu \xi \right) + 2 g_{\mu\nu} \left( \nabla^2\xi \right) \right\}
R^{\mu\nu} h_{\mu\nu} \nonumber \\
&\, + \frac{1}{2}\left\{ \left( - \frac{1}{2\kappa^2} - 4 \nabla^2 \xi \right) \delta^\tau_{\ \mu} \delta^\eta_{\ \nu}
+ 4 \left( \nabla_\rho \nabla_\mu \xi \right) \delta^\eta_{\ \nu} g^{\rho\tau}
+ 4 \left( \nabla_\rho \nabla_\nu \xi \right) \delta^\tau_{\ \mu} g^{\rho\eta}
 - 4g_{\mu\nu} \nabla^\tau \nabla^\eta \xi \right\} \nonumber \\
&\, \qquad \times \left\{ - \nabla^2 h_{\tau\eta} - 2R^{\lambda\ \phi}_{\ \eta\ \tau}h_{\lambda\phi}
+ R^\phi_{\ \tau}h_{\phi\eta} + R^\phi_{\ \tau}h_{\phi\eta} \right\} \nonumber \\
&\, + 2 \left(\nabla^\rho \nabla^\sigma \xi \right)
\left\{ \nabla_\nu \nabla_\rho h_{\sigma\mu}
 - \nabla_\nu \nabla_\mu h_{\sigma\rho}
 - \nabla_\sigma \nabla_\rho h_{\nu\mu}
 + \nabla_\sigma \nabla_\mu h_{\nu\rho}
+ h_{\mu\phi} R^\phi_{\ \rho\nu\sigma}
 - h_{\rho\phi} R^\phi_{\ \mu\nu\sigma} \right\} \nonumber \\
&\, + \frac{1}{2}\frac{\partial T_{\mathrm{matter}\, \mu\nu}}{\partial g_{\tau\eta}}h_{\tau\eta} \, .
\end{align}
The constraint on the propagating speed $c_\mathrm{GW}$ of the gravitational wave is given by the observation of GW170817 
is the following, 
\begin{align}
\label{GWp9} \left| \frac{{c_\mathrm{GW}}^2}{c^2} - 1 \right| < 6
\times 10^{-15}\, .
\end{align}
Here $c$ is the speed of light. 
Let us investigate if the propagating speed $c_\mathrm{GW}$ of the gravitational wave 
$h_{\mu\nu}$ could be equal to or different from that of the light $c$. 
For that, we only need to check the terms including the second derivatives 
of $h_{\mu\nu}$ in (\ref{gb4bD4BoneGW}), 
\begin{align}
\label{second}
I_{\mu\nu} \equiv&\, I^{(1)}_{\mu\nu} + I^{(2)}_{\mu\nu} \, , \nonumber \\
I^{(1)}_{\mu\nu} \equiv&\, \frac{1}{2}\left\{ \left( - \frac{1}{2\kappa^2} - 4 \nabla^2 \xi \right) \delta^\tau_{\ \mu} \delta^\eta_{\ \nu}
+ 4 \left( \nabla_\rho \nabla_\mu \xi\right) \delta^\eta_{\ \nu} g^{\rho\tau}
+ 4 \left( \nabla_\rho \nabla_\nu \xi\right) \delta^\tau_{\ \mu} g^{\rho\eta}
 - 4g_{\mu\nu} \nabla^\tau \nabla^\eta \xi \right\} \nabla^2 h_{\tau\eta} \, , \nonumber \\
I^{(2)}_{\mu\nu} \equiv &\, 2 \left(\nabla^\rho \nabla^\sigma \xi \right)
\left\{ \nabla_\nu \nabla_\rho h_{\sigma\mu}
 - \nabla_\nu \nabla_\mu h_{\sigma\rho}
 - \nabla_\sigma \nabla_\rho h_{\nu\mu}
 + \nabla_\sigma \nabla_\mu h_{\nu\rho} \right\} \, .
\end{align}
Because we assume that the matter minimally couples with gravity, any contribution of the matter does not couple with
any derivative term of $h_{\mu\nu}$, and Eq.~(\ref{second}) does not include any contribution from the matter. 
Therefore matter is irrelevant to the propagating speed of the gravitational wave. 
We should note that $I^{(1)}_{\mu\nu}$ does not change the speed of the gravitational wave from the speed of light.
On the other hand, $I^{(2)}_{\mu\nu}$ may change the speed of the gravitational wave from that of the light and 
as a result, the constraint (\ref{GWp9}) might be violated. 
In the case that $\nabla_\mu \nabla^\nu \xi$ is proportional to the metric $g_{\mu\nu}$ as follows,
\begin{align}
\label{condition}
\nabla_\mu \nabla_\nu \xi = \frac{1}{4}g_{\mu\nu} \nabla^2 \xi \, ,
\end{align}
$I^{(2)}_{\mu\nu}$ does not change the speed of the gravitational wave from that of light, either. 
Note that because $\xi$ is a function specifying the model. Eq.~(\ref{condition}) restricts models 
so that the propagating speed of the gravitational wave coincides with that of light. 
Therefore, we proved that the condition (\ref{condition}) does not change 
from the standard sEGB gravity~\cite{Nojiri:2023jtf, Nojiri:2023mbo}. 

\section{Cosmology\label{SecIV}}

Before considering the spherically symmetric and static spacetime, we may consider the cosmology because 
the constraint (\ref{gfEGB3}) is valid for both the FLRW universe (\ref{FLRW}) and spherically symmetric and static spacetime. 

\subsection{Model construction}

In the FLRW universe (\ref{FLRW}), we find 
\begin{align}
\label{E2}
& \Gamma^t_{ij}= a^2 H \delta_{ij}\, ,\quad \Gamma^i_{jt}=\Gamma^i_{tj}=H\delta^i_{\ j}\, , \nonumber \\
& R_{itjt}= -\left(\dot H + H^2\right)a^2h_{ij}\, ,\quad
R_{ijkl}= a^4 H^2 \left(\delta_{ik} \delta_{lj} - \delta_{il} \delta_{kj}\right)\, ,\nonumber \\
& R_{tt}=-3\left(\dot H + H^2\right)\, ,\quad R_{ij}= a^2
\left(\dot H + 3H^2\right) \delta_{ij}\, ,\quad R= 6\dot H + 12
H^2\, , \nonumber \\
&\, \mbox{other components}=0\, 
\end{align}
We may also assume $\phi$ only depends on the cosmological time $t$. 

By using Eq.~(\ref{gb4bD4one}), we find the equations corresponding to the FLRW equations in the following forms, 
\begin{align}
\label{SEGB3}
0=&\, - \frac{3}{\kappa^2}H^2 + \left( - 2 \lambda \omega(\phi) + \frac{1}{2}\right){\dot\phi}^2 + V(\phi)
+ 24 H^3 \frac{d \xi(\phi(t))}{dt}\, ,\nonumber \\
0=&\, \frac{1}{\kappa^2}\left(2\dot H + 3 H^2 \right) + \frac{1}{2}{\dot\phi}^2 - V(\phi)
 - 8H^2 \frac{d^2 \xi(\phi(t))}{dt^2}
 - 16H \dot H \frac{d\xi(\phi(t))}{dt} - 16 H^3 \frac{d \xi(\phi(t))}{dt} \, .
\end{align}
Here we used the constraint (\ref{gfEGB3}), which has the following form 
\begin{align}
\label{gfEGB3FLRW}
0= \omega(\phi) {\dot\phi}^2 + 1 \, .
\end{align}
In terms of the $e$-foldings number $N$ defined by $a=a_0\e^N$ instead of the cosmological time $t$, Eq.~(\ref{SEGB3}) can be rewritten as follows,
\begin{align}
\label{SEGB3N}
0=&\, - \frac{3}{\kappa^2}H^2 + \left( - 2 \lambda \omega(\phi) + \frac{1}{2}\right) H^2 \phi'(N)^2 + V(\phi) + 24 H^4 \frac{d \xi(\phi(N))}{dN}\, ,\nonumber \\
0=&\, \frac{1}{\kappa^2}\left(2H H' + 3 H^2 \right) + \frac{1}{2}H^2 \phi'(N)^2 - V(\phi)
 - 8H^4 \frac{d^2 \xi(\phi(t))}{dN^2}
 - 24 H^3 \frac{dH}{dN} \frac{d\xi(\phi(N))}{dN} \nonumber \\
&\, - 16 H^4 \frac{d \xi(\phi(t))}{dN} \, . 
\end{align}
Here we used the relations, $\frac{d}{dt}=H \frac{d}{dN}$ and $\frac{d^2}{dt^2}= H^2 \frac{d^2}{dN^2} + \frac{dH}{dN}\frac{d}{dN}$. 

By deleting $V(\phi)$ in (\ref{SEGB3N}), we find 
\begin{align}
\label{GBeq1}
0=&\, \frac{2}{\kappa^2} H'(N) + \left( - 2 \lambda \omega(\phi) + 1 \right) H(N) \phi'(N)^2
 - 8 H(N)^3 \frac{d^2 \xi(\phi(t))}{dN^2} \nonumber \\
&\, - 24 H(N)^2 \frac{dH}{dN} \frac{d\xi(\phi(N))}{dN} +8 H(N)^3 \frac{d \xi(\phi(t))}{dN} \, ,
\end{align}
which can be integrated with respect to $\xi(N)$ and we obtain,
\begin{align}
\label{SEGB10}
\xi(\phi(N))=&\, \frac{1}{8}\int^N dN_1 \frac{\e^{N_1}}{H(N_1)^3} \nonumber \\
&\, \times \int^{N_1} \frac{dN_2}{\e^{N_2}}
\left(\frac{2}{\kappa^2}H' (N_2) + \left( - 2 \lambda \left(N_2\right) \omega\left( \phi \left(N_2\right) \right) + 1 \right) H(N_2) {\phi'(N_2)}^2 \right)\, .
\end{align}
By substituting Eq.~(\ref{SEGB10}) into the first equation in (\ref{SEGB3N}), we find
\begin{align}
\label{SEGB11}
V(\phi(N)) &\, = \frac{3}{\kappa^2}H(N)^2 - \left( - 2 \lambda (N) \omega(\phi) + \frac{1}{2}\right) H(N)^2 \phi' (N)^2 \nonumber \\
& - 3\e^N H(N) \int^N \frac{dN_1}{\e^{N_1}} \left(\frac{2}{\kappa^2} H' (N_1) + \left( - 2 \lambda \left(N_1\right) \omega\left( \phi \left(N_1\right) \right) 
+ 1 \right) H(N_1) \phi'(N_1)^2 \right)\, . \nonumber \\
\end{align}
Just for the simplicity, we choose $\omega(\phi)=-1$ in (\ref{gfEGB3FLRW}), which gives, 
\begin{align}
\label{phidash}
\phi'=\frac{1}{H} \, ,
\end{align}
or $\phi=t$. 
In the case that the $e$-foldings number $N$ is given by $N=f(t)$, we find $N=f(\phi)$ and $H= f' (t)$, 
whcih tells that $\xi(\phi)$ and $V(\phi)$ are given as follows, 
\begin{align}
\label{SEGB12}
V(\phi) =&\, \frac{3}{\kappa^2} \left( f' (\phi)\right)^2
+ 2 \lambda \left( f\left(\phi\right) \right) \omega(\phi) - \frac{1}{2} \nonumber \\
&\, - 3 f'(\phi) \e^{ f(\phi)} \int^\phi d\phi_1 \e^{- f(\phi_1)} \left(\frac{2}{\kappa^2} f''(\phi_1) 
 - 2 \lambda \left( f\left(\phi_1\right) \right) \omega\left( \phi_1 \right) 
+ 1 \right)\, , \\
\label{SEGB13}
\xi(\phi) =&\, \frac{1}{8}\int^\phi d\phi_1
\frac{ \e^{ f(\phi_1)} }{f'(\phi_1)^2}
\int^{\phi_1} d\phi_2 \e^{- f(\phi_2)} \left(\frac{2}{\kappa^2} f''(\phi_2) 
 - 2 \lambda \left(f\left(\phi_2\right)\right) \omega\left( \phi_2 \right) + 1 \right)\, .
\end{align}
These expressions (\ref{SEGB12}) and (\ref{SEGB13}) tell that 
if we consider the model given by (\ref{SEGB12}) and (\ref{SEGB13}), we find one of solution of Eq.~(\ref{SEGB3}) as follows, 
\begin{align}
\label{SEGB14}
\phi= f^{-1}(N)\quad \left(N= f(\phi)\right)\, ,\quad H = f'\left( f^{-1} \left(N\right) \right) \, .
\end{align}
We should note that $\lambda$ can be arbitrary as a function of $t$ or $N$ in (\ref{SEGB12}) and (\ref{SEGB13}) and we may put $\lambda=0$ 
in (\ref{SEGB12}) and (\ref{SEGB13}), 
\begin{align}
\label{SEGB12B}
V(\phi) =&\, \frac{3}{\kappa^2} \left( f' (\phi)\right)^2 - \frac{1}{2} 
 - 3 f'(\phi) \e^{ f(\phi)} \int^\phi d\phi_1 \e^{- f(\phi_1)} \left(\frac{2}{\kappa^2} f''(\phi_1) + 1 \right)\, , \\
\label{SEGB13B}
\xi(\phi) =&\, \frac{1}{8}\int^\phi d\phi_1
\frac{ \e^{ f(\phi_1)} }{f'(\phi_1)^2}
\int^{\phi_1} d\phi_2 \e^{- f(\phi_2)} \left(\frac{2}{\kappa^2} f''(\phi_2) + 1 \right)\, .
\end{align}
Then in the solution of Eq.~(\ref{SEGB3}), $\lambda$ vanishes. 
Therefore for the time-evolution of $H$ given by an arbitrary function $h(N)\equiv f'\left( f^{-1} \left(N\right) \right)$ as in (\ref{SEGB14}), 
we can construct a model realizing the Hubble rate $H$, $H=h(N)$. 

Let us compare the above results with that of sEGB gravity.
In the case of the sEGB gravity in (\ref{I8one}), we have the equations 
corresponding to (\ref{SEGB12B}) and (\ref{SEGB13B}) 
are given by
\begin{align}
\label{SEGB12BB}
V(\phi) =&\, \frac{3}{\kappa^2}h \left( f(\phi)\right)^2
 - \frac{h\left( f\left(\phi\right)\right)^2}{2 f'(\phi)^2} \nonumber \\
&\, - 3h\left( f(\phi)\right) \e^{ f(\phi)}
\int^\phi d\phi_1 f'( \phi_1 ) \e^{- f(\phi_1)} \left(\frac{2}{\kappa^2}h'\left( f(\phi_1)\right)
+ \frac{h'\left( f\left(\phi_1\right)\right)}{ f'(\phi_1 )^2} \right)\, , \\
\label{SEGB13BB}
\xi(\phi) =&\, \frac{1}{8}\int^\phi d\phi_1
\frac{ f'(\phi_1) \e^{ f(\phi_1)} }{h\left( f\left(\phi_1\right)\right)^3}
\int^{\phi_1} d\phi_2 f'(\phi_2) \e^{- f(\phi_2)} \left(\frac{2}{\kappa^2}h'\left( f(\phi_2)\right)
+ \frac{h\left( f(\phi_2)\right)}{ f'(\phi_2)^2} \right)\, ,
\end{align}
which also give the solutions (\ref{SEGB14}). 
Therefore we can construct models in 
both the ghost-free $f(\mathcal{G})$ gravity in (\ref{gfEGB4}) and the scalar--Einstein--Gauss-Bonnet gravity in (\ref{I8one}), 
which realise identical evolution of the expanding universe. 
The difference is, however, that the scalar field does not propagate in the ghost-free $f(\mathcal{G})$ gravity but 
do in the sEGB gravity. 
The difference can be observed in the study of the fluctuations of the universe. 

As a simple example, we consider a model mimicking the $\Lambda$CDM model, 
where the scale factor is given by 
\begin{align}
\label{LCDM1}
a(t) = \e^N = a_0 \sinh^\frac{2}{3} \left( \alpha t \right)\, .
\end{align}
Here $a_0$ and $\alpha$ are positive constants. 
Then we find
\begin{align}
\label{LCDM2}
f(\phi) = \ln \left( a_0 \sinh^\frac{2}{3} \left( \alpha \phi \right) \right) \, .
\end{align}
Then by using (\ref{SEGB12B}) and (\ref{SEGB13B}), we can construct a model reproducing the time evolution of 
the $\Lambda$CDM model in the framework of the ghost-free $f(\mathcal{G})$ gravity in (\ref{gfEGB4}) without real cold dark matter (CDM) which role is taken by non-dynamical scalar. 
If we use (\ref{SEGB12BB}) and (\ref{SEGB13BB}), we obtain a model of the sEGB gravity in (\ref{I8one}) 
realising the identical time evolution of the $\Lambda$CDM model. 
The cold dark matter has no pressure and therefore there is no pressure wave and the fluctuation does not propagate. 
In the case of the sEGB gravity, there appear the propagating scalar model although there is no propagating mode 
in the ghost-free $f(\mathcal{G})$ gravity. 
In this sense, if we consider the perturbation, the ghost-free $f(\mathcal{G})$ gravity behaves more like $\Lambda$CDM model if
compared with the sEGB gravity.

\subsection{A model of inflation}

As another example, we may consider the following as a model of the inflation~\cite{Nojiri:2023mbo}, 
\begin{align}
\label{Ex2}
H=h(N) =H_0 \left( 1 + \alpha N^\beta \right)^\gamma \, .
\end{align}
Because $H \sim H_0 \left( 1 + \alpha \gamma N^\beta \right)$ when $N$ is small 
and$H \sim H_1 \alpha^\gamma N^{\beta\gamma}$ when $N$ is large, 
$H_1$ and $\alpha$ must be positive so that $H$ is positive.
By requiring $\beta>0$ and $\gamma<0$, $H$ becomes a monotonically decreasing function. 

The effective equation of state parameter $w_\mathrm{eff} \equiv -1 - \frac{2\dot H}{3 H^2}$ is given by,
\begin{align}
\label{EC3}
w_\mathrm{eff} = - 1 - \frac{2 \alpha \beta \gamma N^{\beta - 1}}{3 \left( 1 + \alpha N^\beta \right) } \, .
\end{align}
We find that $w_\mathrm{eff}$ goes to $-1$ when $N\to 0$, which could correspond to the early Universe, 
 or $N\to \infty$, which could to the present or future Universe.
Hence the model (\ref{Ex2}) could describe the inflation and the accelerating expansion in the late Universe in a unified way.

We should note that the model (\ref{Ex2}) could also describe the matter-dominated era when $w_\mathrm{eff} \sim 0$. 
If $w_\mathrm{eff} = - \frac{1}{3}$, we obtain 
\begin{align}
\label{EC4}
 - \alpha \beta \gamma N^{\beta - 1} = 1 + \alpha N^\beta \, ,
\end{align}
which has two real and positive solutions. 
We denote the two solutions of (\ref{EC4}) by $N=N_1$ and $N=N_2$ and assume $0<N_1<N_2$.
The period when $N<N_1$ could correspond to the inflation in the early Universe and the period where $N>N_2$ 
to the accelerating expansion in the present Universe. 
During $N_1<N<N_2$, we expect $w_\mathrm{eff}$ goes to vanish at least if we include the contribution of the matter, 
whose equation of state parameter vanishes. 
If $1<\beta<2$ and $\gamma\sim \mathcal{O}\left(10^{12} \right)$, 
the model (\ref{Ex2}) could be realistic as found in \cite{Nojiri:2023mbo}. 

We find $H(N)=f'(\phi)$ because $N=f(\phi)$, and (\ref{Ex2}) tells 
\begin{align}
\label{Hfp1}
H_0 \left( 1 + \alpha N^\beta \right)^\gamma = \frac{dN}{d\phi}\, ,
\end{align}
which gives, 
\begin{align}
\label{Hfp2}
\phi = \int \frac{dN}{H_0 \left( 1 + \alpha N^\beta \right)^\gamma}\, .
\end{align}
We now consider the limit $\beta\to 1$ although $1<\beta<2$ for the realistic model, 
which allows us to integrate Eq.~(\ref{Hfp2}), 
\begin{align}
\label{Hfp3}
\phi = f^{-1}\left(N\right) = \frac{1}{\alpha \left( 1 - \gamma \right) H_0} \left( 1 + \alpha N \right)^{1-\gamma}\, ,
\end{align}
which gives, 
\begin{align}
\label{Hfp4}
N=f(\phi) = \frac{\left\{ \alpha \left( 1 - \gamma \right) H_0 \phi \right\}^\frac{1}{1-\gamma} - 1}{\alpha}\, .
\end{align}
Then Eqs.~(\ref{SEGB12B}) and (\ref{SEGB13B}) gives the corresponding forms of $V(\phi)$ and $\xi(\phi)$. 

The evolution history of the expanding universe identical to (\ref{Ex2}) can be realised also in the framework 
of the standard sEGB gravity. 
The difference appears when we consider the fluctuations because the scalar field is dynamical in sEGB gravity. In our model, the scalar field is not dynamical 
due to the constraint (\ref{gfEGB3}) and there could not be a contribution to the scalar fluctuation. 
This shows that the tensor-to-scalar ratio could be drastically changed in the ghost-free $f(\mathcal{G})$ gravity from that in 
the standard sEGB gravity. 
The fluctuation of the scalar mode is generated by the mixing of the perturbed scalar mode in the metric and any scalar field. 
In the ghost-free $f(\mathcal{G})$, the fluctuation of the scalar field $\phi$ is prohibited by the constraint (\ref{gfEGB3}) 
and therefore the fluctuation of the scalar mode could be suppressed and if there are no other dynamical scalar fields besides $\phi$, 
the tensor-to-scalar ratio might violate the observed constraint. 

\section{Spherically symmetric and static spacetime\label{SecV}}

We now consider the spherically symmetric and static spacetime in (\ref{metD}) by using the model in (\ref{gfEGB4}). 

\subsection{Model construction}

We denote the metric $\tilde g_{ij}$ of the unit sphere by
$d{\Omega_2}^2=\sum_{i,j=1,2} \tilde g_{ij} dx^i dx^j = d\theta^2 + \sin^2\theta d\phi^2$.
For the metric (\ref{metD}), the non-vanishing components of the connections are given as follows, 
\begin{align}
\label{GBv0}
&\Gamma^t_{tt}=\Gamma^t_{rr} = \Gamma^r_{tr} = \Gamma^r_{rt} = 0 \, , \quad \Gamma^r_{tt} = \e^{-2\eta + 2\nu}\nu' \, ,
\quad \Gamma^t_{tr}=\Gamma^t_{rt}=\nu'\, , \quad \Gamma^r_{rr}=\eta'\, ,\nonumber \\
&\Gamma^i_{jk} = \tilde \Gamma^i_{jk}\, ,\quad \Gamma^r_{ij}=-\e^{-2\eta}r\tilde g_{ij} \, ,
\quad \Gamma^i_{rj}=\Gamma^i_{jr}=\frac{1}{r}\delta^i_{\ j}\, .
\end{align}
Here $\tilde \Gamma^i_{jk}$ is the connection defined by $\tilde g_{ij}$.
Because we have 
\begin{equation}
\label{Riemann}
R^\eta_{\ \mu\rho\nu}=
 -\Gamma^\eta_{\mu\rho,\nu}
+ \Gamma^\eta_{\mu\nu,\rho}
 - \Gamma^\eta_{\mu\rho}\Gamma^\eta_{\nu\eta}
+ \Gamma^\eta_{\mu\nu}\Gamma^\eta_{\rho\eta} \, ,
\end{equation}
we also find 
\begin{align}
\label{curvatures}
R_{rtrt} =&\, \e^{2\nu}\left\{\nu'' + \left(\nu' - \eta'\right)\nu' \right\} \, ,\quad
R_{titj} = r\nu'\e^{2(\nu - \eta)} \tilde g_{ij} \, ,\quad 
R_{rirj} = \eta' r \tilde g_{ij} \, ,\nonumber \\
R_{tirj}=&\, 0 \, , \quad
R_{ijkl} = \left( 1 - \e^{-2\eta}\right) r^2
\left(\tilde g_{ik} \tilde g_{jl} - \tilde g_{il} \tilde g_{jk} \right)\, ,\nonumber \\
R_{tt}=&\, \e^{2\left(\nu - \eta\right)} \left\{
\nu'' + \left(\nu' - \eta'\right)\nu' + \frac{2\nu'}{r}\right\} \, , \quad 
R_{rr} = - \left\{ \nu'' + \left(\nu' - \eta'\right)\nu' \right\}
+ \frac{2 \eta'}{r} \, ,\nonumber \\
R_{tr} =&\, 0 \, , \quad
R_{ij} = \left[ 1 + \left\{ - 1 - r \left(\nu' - \eta' \right)\right\}\e^{-2\eta}\right]
\tilde g_{ij}\, , \nonumber \\ 
R=&\, \e^{-2\eta}\left[ - 2\nu'' - 2\left(\nu' - \eta'\right)\nu' - \frac{4\left(\nu' - \eta'\right)}{r} + \frac{2\e^{2\eta} - 2}{r^2} \right] \, .
\end{align}
The metric given by Eq.~(\ref{metD}) gives the $(t,t)$-, $(r,r)$-, and
the angular components of Eq.~(\ref{gb4bD4one}) in the following forms,
\begin{align}
\label{Eq2tt}
 - 4r^2 \e^{2\eta} \kappa^2 \rho =&\,- 16 \left(1 - \e^{-2\eta}\right) \xi'' - 4 \left\{ - 4\left( 1-3\e^{-2\eta} \right)\xi' + r \right\} \eta'
+2 +r^2 \phi'^2 \nonumber \\
&\, + 2 \e^{2\eta} \left( V r^2 -1 \right) \,, \\
\label{Eq2rr}
4 r^2 \e^{2\eta} \kappa^2 p =&\, 4\left\{ - 4 \left(1 -3\e^{-2\eta} \right) \xi' +r \right\} \nu' + 2 - r^2 \phi'^2 -2 \e^{-2\eta} 
+ 2 \e^{2\eta} \left( V + 2 \mu^4 \lambda \right) r^2 \,, \\
\label{Eq2pp}
8r \e^{2\eta} \kappa^2 p =&\, 2 \left(r + 8 \xi' \e^{-2\eta} \right) \left( \nu'' + {\nu'}^2 \right) + 16 \xi'' \nu' \e^{-2\eta}
+ \left\{ -2 \left( r + 24 \xi' \e^{-2\eta} \right) \eta' + 2 \right\} \nu' \nonumber \\
&\, -2 \eta' + r\left( {\phi'}^2 + 2 \e^{2\eta} V \right) \, , 
\end{align}
Here we used (\ref{cnstrnt2}) and $\rho$ is the energy density and $p$ is the pressure of matter, respectively. 
We now assume the matter to be a perfect fluid and it satisfies an equation of state, $p=p\left(\rho\right)$, 
which satisfies the following conservation law,
\begin{align}
\label{FRN2}
0 = \nabla^\mu T_{\mu r} =\nu' \left( \rho + p \right) + \frac{dp}{dr} \, .
\end{align}
Here it has been assumed that $\rho$ and $p$ depend only on the radial coordinate $r$. 
We find other components of the conservation law are trivially satisfied. 
When the equation of state $\rho=\rho(p)$ is given, we can integrate Eq.~(\ref{FRN2}) as follows,
\begin{align}
\label{FRN3}
\nu = - \int^r dr \frac{\frac{dp}{dr}}{\rho + p}
= - \int^{p(r)}\frac{dp}{\rho(p) + p} \, .
\end{align}
On the other hand, Eq.~(\ref{Eq2tt}) tells, 
\begin{align}
\label{V}
V=&\, - 2 \kappa^2 \rho + \frac{1}{r^2} \left\{ 
8 \e^{-2\eta} \left(1 - \e^{-2\eta}\right) \xi'' + 2 \e^{-2\eta} \left\{ - 4\left( 1-3\e^{-2\eta} \right)\xi' + r \right\} \eta' 
 - \e^{-2\eta} \right\} \nonumber \\
&\, - \frac{1}{2} \e^{-2\eta} \phi'^2 \, .
\end{align} 
By combining Eq.~(\ref{Eq2tt}) with Eq.~(\ref{Eq2rr}), we find, 
\begin{align}
\label{lambda}
\mu^4 \lambda=&\, \kappa^2 \left(\rho + p\right) 
+ \frac{\e^{-2\eta}}{r^2} \left[ - 4 \left(1 - \e^{-2\eta}\right) \xi'' - \left\{ - 4\left( 1-3\e^{-2\eta} \right)\xi' + r \right\} \eta' 
\right. \nonumber \\
&\, \left. - \left\{ - 4 \left(1 -3\e^{-2\eta} \right) \xi' +r \right\} \nu' \right] 
+ \frac{\e^{-2\eta}}{2r^2} \left( r^2 \phi'^2 + \e^{-2\eta} - \e^{2\eta} \right) \, .
\end{align}
The combination of Eq.~(\ref{Eq2tt}) and Eq.~(\ref{Eq2pp}) also gives,
\begin{align}
\label{f2}
0 =&\, - 8\, \left\{ \e^{-2\eta} \left( \nu' r - 1 \right) +1 \right\} \xi'' - 8 \e^{-2\eta} \left\{ r \left( \nu'' + {\nu'}^2 -3 \nu' \eta' \right)
+ \eta' \left( 3 -\e^{2\eta} \right) \right\} \xi' \nonumber \\
&\, -r^2 \left( \nu'' +{\nu'}^2 -\nu' \eta' \right) -2r \left( \nu' +\eta' \right) - \e^{2\eta} +1 - 2 \kappa^2 r^2 \e^{2\eta} \left( \rho + p \right)\, .
\end{align}
The equation~(\ref{f2}) is regarded as a differential equation for $\xi'$ and therefore for $\xi$. 
If $\nu=\nu(r)$, $\eta = \eta(r)$, $\rho=\rho(r)$, and $p=p(r)$ are given, the solution is,
\begin{align}
\label{f3}
\xi(r) = &\, - \frac{1}{8}\int \Bigl[ \int \frac{
\e^{2\eta} \left\{ \e^{2\eta} + r^2 \left( \nu'' +{\nu'}^2 -\nu' \eta' \right) +r (\nu' +\eta') -1 - 2 \kappa^2 r^2 \e^{2\eta} \left( \rho + p \right)\right\} }
{U \left(\nu' r - 1 + \e^{2\eta} \right)} dr \nonumber \\
&\, + c_1 \Bigr] U dr +c_2 \, , \nonumber \\
U(r) \equiv&\, \exp \left\{ -\int \frac{ r \left( \nu'' + {\nu'}^2 \right) + \eta' \left( 3 - \e^{2\eta} - 3 \nu' r
\right)} {\nu' r - 1 + \e^{2\eta}} dr \right\} \, ,
\end{align}
where $c_1$ and $c_2$ are constants of integration. 
When we properly assume the profiles of $\nu=\nu(r)$ and $\eta = \eta(r)$, 
by using (\ref{f3}), we find the $r$-dependence of $\xi$, $\xi=\xi(r)$. 
On the other hand, by solving (\ref{cnstrnt2}), we find the $r$-dependence of $\phi$, $\phi=\phi(r)$, which could be solved with respect to $r$, $r=r(\phi)$. 
Then by using (\ref{f3}), we can find $\xi$ as a function of $\phi$, $\xi=\xi(\phi)$. 
Furthermore, by using (\ref{V}), we find $V$ as a function of $\phi$, $V=V(\phi)$. 
On the other hand, Eq.~(\ref{lambda}) gives $\lambda$ as a function of $r$, which is a solution of the model. 

In the case that $\e^{2\nu}=\e^{-2\eta}$ and there are no contributions from matter, $\rho=p=0$ as in the Schwarzschild spacetime, 
by using Eqs.~(\ref{V}), (\ref{lambda}), and (\ref{f3}), we find 
\begin{align}
\label{VHay}
V=&\, \frac{1}{r^2} \left\{ 
8 \e^{-2\eta} \left(1 - \e^{2\nu}\right) \xi'' + 2 \e^{2\nu} \left\{ 4\left( 1-3\e^{2\nu} \right)\xi' + r \right\} \nu'
 - \e^{2\nu} \right\}- \frac{1}{2} \e^{2\nu} \phi'^2 \, , \\
\label{lambdaHay}
\mu^4 \lambda=&\, \frac{\e^{2\nu}}{r^2} \left\{ - 4 \left(1 - \e^{2\nu}\right) \xi'' - 8 \left( 1-3\e^{2\nu} \right)\xi' \nu' \right\} 
+ \frac{\e^{2\nu} \phi'^2 }{2} \, , \\
\label{f3Hay}
U(r) =&\, \exp \left\{ -\int \frac{ r \left( \nu'' + 4 {\nu'}^2 \right) \e^{4\nu} - 3 \nu' \e^{4\nu} + \nu' \e^{2\nu} }
{\left(\nu' r - 1 \right) \e^{4\nu} + \e^{2\nu}} dr \right\} \, , \nonumber \\
\xi(r) =&\, - \frac{1}{8}\int \left[ \int \frac{
1 + \left\{ r^2 \left( \nu'' + {\nu'}^2 \right) -1\right\} \e^{2\nu} } 
{U \left\{ \left(\nu' r - 1 \right) \e^{4\nu} + \e^{2\nu} \right\}} dr + c_1 \right] U dr +c_2 \, .
\end{align}
The above expressions are used below. 

Let us consider whether there are any other solutions besides the assumed profiles of $\nu=\nu(r)$ and $\eta = \eta(r)$. 
For this purpose, we assume that the given spacetime is asymptotically flat and 
we show that any solution in Einstein's gravity without cosmological constant, with or without matter, is also a solution to the model.

For simplicity, we assume $\omega(\phi)$ is given by (\ref{ex4}). 
The assumption that the spacetime is asymptotically flat when $r \rightarrow \infty$ tells that
the flat spacetime is also a solution automatically 
if we consider the limit $r_0 \rightarrow \infty$ after shifting the radial coordinate $r \rightarrow r_0+r$. 
Here $\lambda \omega(\phi) - \frac{1}{2}$ goes to vanish. 
Although the scalar field $\phi=\mu^2 \left( r_0+r\right)$ goes to $\infty$ in these limits, we can redefine the scalar field to
remain finite, for example, by using the redefinition $\Phi={\phi}^{-1}$. 
We should note that in the obtained solution describing flat spacetime, 
the scalar fields $\phi$ or $\Phi$ cannot be identified with the radial coordinate $r$ but it is constant $\Phi=0$ everywhere in the spacetime. 

In the limit $\phi \rightarrow \infty$, $\lambda \mu^4 + V(\phi)$ vanishes and $\xi(\phi)$ becomes a constant. 
In terms of $\Phi$, the extremum (minimum or maximum) of the scalar field potential
$V\left(\phi=\Phi^{-1} \right)$ is given by $\Phi=0$. 
This tells why flat spacetime is a solution. 
We should also note that in the solution where $\Phi=0$ and $\lambda=0$, 
the field equation~(\ref{I10one}) is satisfied everywhere in the spacetime. 
Furthermore, the equation corresponding to the Einstein equation in (\ref{gb4bD4one}), 
reduces to the standard Einstein equation without the scalar field $\phi$, 
\begin{align}
\label{gb4bD4moddec0}
0= \frac{1}{2\kappa^2}\left(- R_{\mu\nu} + \frac{1}{2} g_{\mu\nu} R\right) 
+ \frac{1}{2} T_{\mathrm{matter}\, \mu\nu} \, .
\end{align}
Therefore, not only a flat spacetime but the standard cosmological solutions like the Schwarzschild black hole, 
Kerr black hole, etc. or self-gravitating objects like standard stellar solutions including neutron stars, etc. in Einstein's gravity 
are surely also solutions in this model. 

For sEGB gravity in (\ref{I8one}), instead of (\ref{VHay}) and (\ref{f3Hay}), we find the following expressions, 
\begin{align}
\label{V2GB}
V =&\, \kappa^2 \left( - \rho + p \right) + \frac{\e^{-2\lambda}}{r^2}\left\{ - 4 \left( \e^{-2\lambda}-1 \right) \xi'' - 4 \left( 1-3\e^{-2\lambda} \right) (\lambda' -\nu')
 \xi' +\e^{2\lambda} - 1 \right\} \nonumber \\
&\, +\frac{\e^{-2\lambda}}{r} \left( \lambda' -\nu' \right) \, , \\
\label{f3sGB}
\xi(r) = &\, - \frac{1}{8}\int \left[ \int \frac{
\e^{2\lambda} \left\{ \e^{2\lambda} + r^2 \left( \nu'' +{\nu'}^2 -\nu' \lambda' \right) +r (\nu' +\lambda') -1 \right\} }
{U \left(\nu' r - 1 + \e^{2\lambda} \right)} dr + c_1 \right] U dr +c_2
 \, , \nonumber \\
U(r) \equiv&\, \exp \left\{ -\int \frac{ r \left( \nu'' +
{\nu'}^2 \right) + \lambda' \left( 3 - \e^{2\lambda} - 3 \nu' r
\right)} {\nu' r - 1 + \e^{2\lambda}} dr \right\} \, ,
\end{align}
where $c_1$ and $c_2$ are integration constants, again. 
The scalar field $\phi$ is given as a function of $r$ as follows, 
\begin{align}
\label{phiGB}
\phi = \pm \int \sqrt{ - \frac{8}{r^2} \left\{ \left( \e^{-2\lambda}-1 \right) \xi''
+ \left( 1-3\e^{-2\lambda} \right) \left( \lambda' +\nu' \right) \xi'\right\}
+\frac{2}{r} \left( \lambda' +\nu' \right) } dr \, .
\end{align}
Then by solving (\ref{phiGB}) with respect $r$, the radial coordinate $r$ becomes a function of the scalar field $\phi$. $r=r(\phi)$. 
By substituting the expression of $r(\phi)$ into the equations in (\ref{V2GB}) and (\ref{f3sGB}), we obtain $V$ and $\xi$ as functions of $\phi$, 
$V=V(\phi)$ and $\xi=\xi(\phi)$. 

In the case of sEGB gravity, however, the expression (\ref{phiGB}) tells that 
in order to avoid that the scalar field $\phi$ could become the ghost or for the scalar field $\phi$ to become a real field, 
we need to require, 
\begin{align}
\label{cons3}
 - \frac{8}{r^2} \left\{ \left( \e^{-2\lambda}-1 \right) \xi''
 + \left( 1-3\e^{-2\lambda} \right) \left( \lambda' +\nu' \right) \xi'\right\}
+\frac{2}{r} \left( \lambda' +\nu' \right) \geq 0 \, .
\end{align}
Therefore the class of the realised geometry is restricted. 
In the case of the ghost-free $f(\mathcal{G})$ gravity, because the scalar field $\phi$ does not propagate, the scalar field cannot be ghost. 

Even if the identical geometry can be realised both in the sEGB gravity and the ghost-free $f(\mathcal{G})$ gravity, 
there could be a difference when we consider the phenomena like the black hole merger. 
Both of the geometry could have a hair of the scalar field. 
When two black holes or steller objects merge, in the case of the scalar--Einstein--Gauss-Bonnet gravity, the scalar wave makes a flux of the energy. 
On the other hand, in the case of the ghost-free $f(\mathcal{G})$ gravity, there is no scalar wave bringing the energy. 
Therefore the flux of the gravitational wave becomes larger in the ghost-free $f(\mathcal{G})$ gravity compared with the sEGB gravity, 
which might be found in the future observations. 


\subsection{Schwarzschild spacetime}\label{Schw}

The Schwarzschild spacetime $\e^{2\eta}= \frac{1}{1 - \frac{r_0}{r}}$ is, of course, a solution of the model. 
We now check how the constraint.~(\ref{cnstrnt2}) works in the Schwarzschild spacetime. 
Eq.~(\ref{cnstrnt2}) has the following form, 
\begin{align}
\label{eq4B1NGB}
\frac{1}{1 - \frac{r_0}{r}} \omega (\phi) \left( \phi' \right)^2 = \mu^4 \, .
\end{align}
We now choose $\omega$ by (\ref{ex1}). 
Then when $\frac{r_0}{r}<1$, Eq.~(\ref{cnstrnt2}) has the following form 
\begin{align}
\label{omo1NGB}
\left( \phi^\frac{1}{2} \right)' = \frac{\mu^2}{4} \sqrt{ 1 - \frac{r_0}{r}} \, .
\end{align}
Therefore 
\begin{align}
\label{omo1NGB2}
\phi^\frac{1}{2} = \frac{r_0 \mu^2}{8} \left\{ \frac{1}{1 - \sqrt{ 1 - \frac{r_0}{r}}} - \frac{1}{1 + \sqrt{ 1 - \frac{r_0}{r}}} 
+ \ln \frac{1 - \sqrt{ 1 - \frac{r_0}{r}}}{1 + \sqrt{ 1 - \frac{r_0}{r}}} \right\} + C_0
\end{align}
Let us choose the constant of the integration $C_0=0$ so that $\phi$ vanishes at the horizon $r=r_0$. 
On the other hand when $\frac{r_0}{r}>1$, Eq.~(\ref{eq4B1NGB}) has the following form 
\begin{align}
\label{omo1BNGB}
\left( \left( - \phi\right)^\frac{1}{2} \right)' = \frac{\mu^2}{4} \sqrt{\frac{r_0}{r} - 1} \, ,
\end{align}
and we obtain 
\begin{align}
\label{omo1B}
\left( - \phi\right)^\frac{1}{2} = \frac{r_0 \mu^2}{8} \left\{ \frac{i}{1 - i \sqrt{ \frac{r_0}{r} - 1}} - \frac{i}{1 + i\sqrt{\frac{r_0}{r}-1}} 
 - i \ln \frac{1 - i\sqrt{ \frac{r_0}{r} - 1}}{1 +i \sqrt{\frac{r_0}{r} - 1}} \right\} + {\tilde C}_0
\end{align}
We choose the constant of the integration ${\tilde C}_0=0$ so that $\phi$ vanishes at the horizon $r=r_0$, again. 

\subsection{Reissner-Nordstr\"{o}m Black Hole} 

We may also consider the spacetime identical to the Reissner-Nordstr\"{o}m black hole, whose metric is given by, 
\begin{align}
\label{RN}
\e^{2\nu} = \e^{-2\eta} =1 - \frac{2M}{r} + \frac{Q^2}{r^2} \, .
\end{align}
Note that there are no electromagnetic fields in our model. 
Therefore the charge $Q$ is not a real electric charge but it appears effectively by the scalar hair in our model. 
Because the model has no electromagnetic gauge symmetry, there is no reason that the effective charge $Q$ is conserved. 

For the metric, the equations in (\ref{f3Hay}) give, 
\begin{align}
\label{UciRN}
U(r) =&\, \frac{r^4}{3M}
\left( r - r_+ \right)^{ \frac{r_+ \left(r_+ - A\right)}{3\left(r_+ - r_-\right) \left(r_+ - B\right)} -1}
\left( r - r_- \right)^{ - \frac{r_- \left(r_- - A\right)}{3\left(r_+ - r_-\right) \left(r_- - B\right) } -1}
\left( r - B \right)^{- \frac{\left(B-A\right) B}{3\left( r_+ - B \right) \left( r_- - B \right)} -1} \, , \nonumber \\
\xi(r) = &\, - \frac{1}{8}\int \left[ \frac{2Q^2 - M^2}{3M}\int \frac{r^2 \left( r -C_+ \right) \left( r - C_- \right) }
{U\left( r - r_+ \right)^2 \left( r - r_- \right)^2 \left( r - B \right)}dr 
+ c_1 \right] U dr +c_2 , . 
\end{align}
Here $r_+$ and $r_-$ are the radius of the outer horizon and inner horizon, given by
\begin{align}
\label{rpm}
r_\pm \equiv M \pm \sqrt{M^2 - Q^2} \, , 
\end{align}
and $A$, $B$, and $C_\pm$ are defined by 
\begin{align}
\label{ABCRN}
A\equiv\frac{Q^2}{M}= \frac{2r_+ r_-}{r_+ + r_-}\, , \quad B\equiv \frac{2}{3}A\, , \quad 
C_\pm \equiv&\, \frac{MQ^2 \pm Q^2 \sqrt{2 \left( M^2 - Q^2 \right)}}{2Q^2 - M^2} \, .
\end{align}
Eqs.~(\ref{VHay}) and (\ref{lambdaHay}) give the forms of the potential $V=V(\phi)$ and the multiplier field $\lambda=\lambda(r)$. 

Scalarisation in gravity theories is the phenomenon where a black hole or compact star which is originally a solution in general relativity, develops 
to obtain a non-trivial scalar hair. 
The spacetime identical to the Reissner-Nordstr\"{o}m black hole in (\ref{RN}) could be regarded as an example of scalarisation because the hair is 
not an electromagnetic field but a scalar field. 
Especially an interesting phenomenon is spontaneous scalarisation, which is a phase transition. 
In spontaneous scalarisation, the configuration with a trivial scalar field becomes that with a non-trivial scalar field. 
It could be possible to consider the phase transition in the configuration of (\ref{RN}) from the viewpoint of thermodynamics. 

\subsection{Hayward black hole}\label{SecIVC}

As another example, we may consider the Hayward black hole~\cite{Hayward:2005gi}, 
\begin{align}
\label{Hay1}
\e^{2\nu}=\e^{-2\eta}= 1 - \frac{r_0 r^2}{r^3 + r_0 \lambda^2}\, .
\end{align}
Here $\lambda$ is a parameter with the dimension of the length and $M=\frac{r_0}{2}$ corresponds to the Arnowitt-Deser-Misner (ADM) mass. 
The Hayward black hole is regular and there is no curvature singularity. 
\begin{enumerate}
\item If $\frac{2^\frac{2}{3}r_0}{3\left( r_0 \lambda^2 \right)^\frac{1}{3}}<1$, 
$\e^{2\nu}$ does not vanish and is positive and the spacetime given by (\ref{Hay1}) is a kind of the gravasar~\cite{Mazur:2001fv}. 
\item When $\frac{2^\frac{2}{3}r_0}{3\left( r_0 \lambda^2 \right)^\frac{1}{3}}>1$, 
$\e^{2\nu}$ vanishes twice corresponding to the outer and inner horizons. 
\item In the case $\frac{2^\frac{5}{3}M}{3\left( 2M\lambda^2 \right)^\frac{1}{3}}=1$, 
the radii of the two horizons coincide with each other corresponding to the extremal black hole. 
\end{enumerate}

Because further calculations are tedious and the explicit results are not simple, we do not give the explicit results. 
If we choose (\ref{ex4}), we obtain (\ref{ex5}), that is, $\phi = r$. 
Then by using (\ref{V}), we find the form of $V=V(\phi)$. 

In \cite{Nojiri:2023qgd}, the model whose solutions include the Hayward black hole has been given 
in the framework of sEGB gravity in (\ref{I8one}) 
but it was difficult to show completely the absence of the ghosts although there is some circumstantial evidence. 
There remains some possibility that the scalar field $\phi$ becomes ghost. 
In the model of the ghost-free $f(\mathcal{G})$ gravity (\ref{gfEGB4}) in this paper, however, any ghosts do not appear 
because the scalar field $\phi$ is not dynamical and does not propagate. 

Even in the Hayward black hole (\ref{Hay1}), the solution has a scalar hair. 
Therefore when we consider the black hole merger or any other violent collision, the scalar field carries the energy 
in the case of sEGB gravity but it does not in the case of the ghost-free $f(\mathcal{G})$ gravity. 
This could tell that the flux of the gravitational wave could be stronger for the ghost-free $f(\mathcal{G})$ gravity than for the sEGB gravity. 

\section{Photon Sphere and Black Hole Shadow\label{SecVI}}

The constraint (\ref{gfEGB2}) is very similar to that appeared in the mimetic gravity, 
where the scalar field called the mimetic scalar appears. 
Recently in remarkable paper \cite{Khodadi:2024ubi}, it was shown that 
the Event Horizon Telescope observations~\cite{EventHorizonTelescope:2019dse} rule out the compact objects in simplest mimetic gravity. 
This work was based on the investigation of the black hole shadow~\cite{Held:2019xde, Perlick:2021aok, Chen:2022scf}. 
Based on the new mimetic constraint similar to (\ref{gfEGB3}), in \cite{Nojiri:2024txy}, it has been shown that the black hole shadow could become 
consistent with the Event Horizon Telescope observations for the improved version of mimetic gravity. 

The radius $r_\mathrm{sh}$ of the black hole shadow is given by the radius $r_\mathrm{ph}$ of the circular orbit of the photon, which is called a photon sphere, as follows, 
\begin{align}
\label{shph}
r_\mathrm{sh}=\left. r\e^{-\nu(r)} \right|_{r=r_\mathrm{ph}}\, .
\end{align}
The motion of the photon is governed by the following Lagrangian, 
\begin{align}
\label{ph1g}
\mathcal{L}= \frac{1}{2} g_{\mu\nu} \dot q^\mu \dot q^\nu = \frac{1}{2} \left( - \e^{2\nu} {\dot t}^2 
+ \e^{2\eta} {\dot r}^2 + r^2 {\dot\theta}^2 + r^2 \sin^2 \theta {\dot\phi}^2 \right) \, .
\end{align}
We express the derivative with respect to the affine parameter by the ``dot'' or ``$\dot\ $''. 
We also require $\mathcal{L}=0$ for the case of a photon, whose geodesic is null. 
Because the Lagrangian $\mathcal{L}$ does not depend on $t$ and $\phi$ explicitly, 
there appear conserved quantities corresponding to energy $E$ and angular momentum $L$, 
\begin{align}
\label{phEg}
E \equiv&\, \frac{\partial \mathcal{L}}{\partial \dot t} = - \e^{2\nu} \dot t \, , \\
\label{phMg}
L \equiv&\, \frac{\partial V}{\partial\dot\phi}= r^2 \sin^2 \theta \dot\phi \, , 
\end{align}
It should be also noted that the total energy $\mathcal{E}$ of the system should be conserved, 
\begin{align}
\label{totalEg}
\mathcal{E} \equiv \mathcal{L} - \dot t \frac{\partial \mathcal{L}}{\partial \dot t} - \dot r \frac{\partial \mathcal{L}}{\partial \dot r} 
 - \dot\theta \frac{\partial \mathcal{L}}{\partial \dot\theta} - \dot\phi \frac{\partial \mathcal{L}}{\partial \dot\phi} = \mathcal{L} \, , 
\end{align}
For the null geodesic, we find $\mathcal{E}=\mathcal{L}=0$. 
Without any loss of generality, we may consider the orbit on the equatorial plane with $\theta=\frac{\pi}{2}$, 
which allows the condition $\mathcal{E}=\mathcal{L}=0$ to give, 
\begin{align}
\label{geo1g}
0= - \frac{E^2}{2} \e^{-2 \left( \nu + \eta\right)} + \frac{1}{2} {\dot r}^2 + \frac{L^2 \e^{- 2\eta}}{2r^2} \, ,
\end{align}
This system is analogous to the classical dynamical system with potential $W(r)$, 
\begin{align}
\label{geo2g}
0 =\frac{1}{2} {\dot r}^2 + W(r)\, , \quad W(r) \equiv \frac{L^2 \e^{- 2\eta}}{2r^2} - \frac{E^2}{2} \e^{-2 \left( \nu + \eta\right)}\, .
\end{align}
The radius of the circular orbit, where $\dot r=0$, is given by $W(r)= W'(r)=0$ by the analogy of classical mechanics. 

For simplicity, we consider the case $\e^{2\eta}=\e^{-2\nu}$ as in the Schwarzschild spacetime. 
Then the solution of $W'(r)=0$ corresponds to the extremum of $\frac{\e^{2\nu}}{r^2}$ 
and the equation $W(r)=0$ becomes the equation determining $E$. 
Then for example, if we consider the spacetime where
\begin{align}
\label{phs}
\frac{\e^{2\nu}}{r^2} = \frac{\left( r - r_\mathrm{ph} \right)^{3n -2} - \left( r_\mathrm{h} - r_\mathrm{ph} \right)^{3n -2}
 - 2 \left( M - r_\mathrm{ph} \right) \left\{ \left( r - r_\mathrm{ph} \right)^{3n - 3} - \left( r_\mathrm{h} - r_\mathrm{ph} \right)^{3n - 3} \right\}}
{\left\{\left( r - r_\mathrm{ph} \right)^n - \left( - r_\mathrm{ph} \right)^n\right\}^3}\, ,
\end{align}
we find $r=r_{\mathrm{ph}}$ is the radius of the photon sphere and $r=r_\mathrm{h}$ is the horizon radius. 
Here $n$ is a constant equal to or larger than $2$, $n\geq 2$. 
The parameter $M$ is also a constant. 
Because $\e^{-2\nu}$ behaves as $\e^{-2\nu}\sim 1 - \frac{2M}{r}$ when $r$ is large, $M$ can be identified with the ADM mass. 
When $r\to 0$, we find 
$\e^{2\nu}\to \frac{ - r_\mathrm{ph} + r_\mathrm{ph} \left( 1 - \frac{r_\mathrm{h}}{r_\mathrm{ph}} \right)^{3n -2}
 - 2 \left( M - r_\mathrm{ph} \right) \left\{ 1 - \left( 1 - \frac{r_\mathrm{h}}{r_\mathrm{ph}} \right)^{3n - 3} \right\}}{n^3 r} $. 
Therefore in the model (\ref{phs}), we can choose, the radius of the photon sphere $r_\mathrm{ph}$, the horizon radius $r_\mathrm{h}$, and the ADM mass $M$ 
as independent parameters. 
We can construct the model corresponding to (\ref{phs}), again, although the explicit form is very complicated. 

Eq.~(\ref{shph}) tells that the radius of the black hole shadow is given by 
\begin{align}
\label{shph2}
r_\mathrm{sh} 
= \left[ \frac{ {r_\mathrm{ph}}^{3n}}
{ \left( r_\mathrm{ph} - r_\mathrm{h} \right)^{3n -2}
 - 2 \left( r_\mathrm{ph} - M \right) \left( r_\mathrm{ph} - r_\mathrm{h} \right)^{3n - 3}}
\right]^\frac{1}{2} \, .
\end{align}
We can also choose the value of $r_\mathrm{sh}$ by adjusting the value of $r_\mathrm{ph}$ so that the model does not conflict with the observation. 
In Ref.~\cite{Bambi:2019tjh}, it has been shown that for M87$^*$, the radius of the black hole shadow is limited to be $2r_\mathrm{sh}/M \sim 11.0\pm 1.5$ and in \cite{Vagnozzi:2022moj}, 
for Sgr A$^*$, $4.21\lesssim r_\mathrm{sh}/M \lesssim 5.56$. 
In Eq.~(\ref{shph2}), these bounds can be satisfied by adjusting the parameters $r_\mathrm{ph}$ and $r_\mathrm{h}$, as follows. 
We may write $r_\mathrm{ph}=a_\mathrm{ph} M$ and $r_\mathrm{h}=a_\mathrm{h} M$. 
Then Eq.~(\ref{shph2}) can be rewritten as 
\begin{align}
\label{shph2a}
\frac{r_\mathrm{sh}}{M} 
= \left[ \frac{ {a_\mathrm{ph}}^{3n}}
{ \left( a_\mathrm{ph} - a_\mathrm{h} \right)^{3n -2}
 - 2 \left( a_\mathrm{ph} - 1 \right) \left( a_\mathrm{ph} - a_\mathrm{h} \right)^{3n - 3}}\right]^\frac{1}{2} \, .
\end{align}
Just for simplicity and demonstration, we may choose $a_\mathrm{ph} = 1$. 
Then for M87$^*$, we find $a_\mathrm{h} \sim 1 - \left( 5.5 \right)^{-\frac{2}{3n-2}}$ amd 
for Sgr A$^*$, $1 - \left( 4.21 \right)^{-\frac{2}{3n-2}} \lesssim a_\mathrm{h} \lesssim 1 - \left( 5.56 \right)^{-\frac{2}{3n-2}}$. 

Because the radius of the photon sphere $r_\mathrm{ph}$, the horizon radius $r_\mathrm{h}$, and the ADM mass $M$ are independent parameters in our model, 
we may consider some exotic cases. 
We should note, however, that at least the radius of the photon sphere cannot be smaller than the horizon radius $r_\mathrm{ph} \geq r_\mathrm{h}$ 
because the photon must fall inside the black hole and therefore there is no circular orbit. 
Far observers could observe the ADM mass $M$ and they might think that the horizon radius could be given by $2M$. 
In our model, we may choose the radius of the photon sphere $r_\mathrm{ph}$ to be smaller than the false horizon radius $2M$, $r_\mathrm{ph}<2M$. 
The radius of the black hole shadow also might be smaller than $2M$, if we choose
\begin{align}
\label{exotics}
\frac{ {r_\mathrm{ph}}^{3n}}
{ \left( r_\mathrm{ph} - r_\mathrm{h} \right)^{3n -2}
 - 2 \left( r_\mathrm{ph} - M \right) \left( r_\mathrm{ph} - r_\mathrm{h} \right)^{3n - 3}} < 4 M^2 \, .
\end{align}
It could be interesting if such exotic objects could be found by any observation. 

The orbit of the photon is determined by the geodesics and therefore 
the radii of the photon sphere and black hole shadow only depend on the geometry of the spacetime 
and they are independent of the gravity theory. 
Then for the case of the Schwarzschild black hole in Subsection~\ref{Schw}, we have $\frac{r_\mathrm{sh}}{M} = 3 \sqrt{3}$. 
In the case of the Reissner-Nordstr\"{o}m black hole in (\ref{RN}), the potential $W(r)$ in (\ref{geo2g}) is given by 
\begin{align}
\label{WRN1}
W(r) = \frac{L^2}{2} \left( \frac{1}{r^2} - \frac{2M}{r^3} + \frac{Q^2}{r^4} \right) - \frac{E^2}{2} \, ,
\end{align}
which gives
\begin{align}
\label{WRN2}
W'(r) = \frac{L^2}{2} \left( - \frac{2}{r^3} + \frac{6M}{r^4} - \frac{4Q^2}{r^5} \right) \, .
\end{align}
Therefore the radius $r_\mathrm{ph}$ of the photon sphere, which is a solution of the equation $W'(r)=0$, is given by 
\begin{align}
\label{phRN1}
r_\mathrm{ph}=\frac{1}{2} \left( 3M + \sqrt{9M^2 - 8Q^2} \right )\, . 
\end{align}
By using (\ref{shph}), the radius of the shadow is estimated to be 
\begin{align}
\label{shphRN}
r_\mathrm{sh} = \frac{9M^2 - 4Q^2 + 3M \sqrt{9M^2 - 8Q^2} }
{\sqrt{ 6 M^2 - 4 Q^2 + 2M \sqrt{9M^2 - 8Q^2} }} \, .
\end{align}
Even for the Hayward black hole~(\ref{Hay1}), we may similarly find the radii of the photon sphere and the black hole shadow. 

The orbit of the photon is determined only by the geometry and therefore it does not depend on the gravity theory. 
Of course, if there is a coupling between the scalar field and the electromagnetic field like, 
\begin{align}
\label{nonminimal}
S_\mathrm{EM} = - \frac{1}{4} \int d^4 x \sqrt{-g} \Phi(\phi) F_{\mu\nu} F^{\mu\nu}\, .
\end{align}
the propagation of the photon could be changed. 
Here $F_{\mu\nu}$ is the field strength of the electromagnetic field and $\Phi(\phi)$ is a function of $\phi$ expressing the coupling 
between the scalar field $\phi$ and the electromagnetic field. 
This situation is not so changed in between the ghost-free $f(\mathcal{G})$ gravity in this paper 
and the standard sEGB gravity. 
If we consider the quantum field theory, however, in the case of the scalar-Gauss-Bonnet model, the scalar particle, 
which is a quantum scalar, might decay into the photon(s) and two or more photons might decay into the scalar particle via 
the interaction described in (\ref{nonminimal}). 
In the case of the ghost-free $f(\mathcal{G})$ gravity, however, because the scalar field does not propagate, such decays are prohibited. 
Therefore in the case of the black hole and neutron star merger, the emission of the photons could be different in the ghost-free $f(\mathcal{G})$ gravity 
and the standard sEGB gravity.

\section{Summary and Discussions\label{SecVII}}

The constraint (\ref{gfEGB2}), which appears in the ghost-free $f(\mathcal{G})$ gravity model in \cite{Nojiri:2018ouv} has difficulties when we consider 
the spacetime with the horizon(s) or when we consider the expanding universe with account of the horizon.
In the case of a similar constraint that appeared in the mimetic gravity theory, the problem has been solved by modifying the constraint as in (\ref{gfEGB3})~\cite{Nojiri:2022cah}. 
In this paper, the improved formulation was also applied to the ghost-free $f(\mathcal{G})$ gravity model, which allows us 
to discuss the spacetime of the expanding universe and static black hole spacetime including the problem of the black hole shadow. 

By using the obtained model, we discussed the gravitational wave and found the condition that the propagating speed of the gravitational wave 
is identical to that of the propagation speed of light although it is identical to that found in \cite{Nojiri:2023jtf, Nojiri:2023mbo} for sEGB gravity. 
We also proposed a model that describes inflation in the early universe and could satisfy the observational constraints 
while, unlike the sEGB gravity, the non-dynamical scalar may play the role of CDM. 
It has also been discussed how we obtain a model realising the given static spacetime with spherical symmetry. 
Especially the models realising the Reissner-Nordstr\"{o}m black hole and the Hayward black hole, which is a regular black hole without curvature singularity, 
were proposed. 
We also proposed a model which includes the Arnowitt-Deser-Misner (ADM) mass, the horizon radius, and the radius of the photon sphere 
as independent parameters and discussed the radius of the black hole shadow in this model. It is shown that there is a parameter region 
which satisfies the constraints coming from the observations so that the theory under consideration is consistent 
with the Event Horizon Telescope observations~\cite{EventHorizonTelescope:2019dse}

It could be interesting to consider other kinds of geometries like neutron stars and wormholes for ghost-free gravity under consideration.
For example, in \cite{Nojiri:2023ztz}, compact stellar objects whose radius is smaller than the Schwarzschild radius defined by ADM mass were 
constructed. 
In the framework of the ghost-free version proposed in this paper, we may consider  similar problems as well as dynamical wormhole solutions. Moreover, one can generalise the study of primordial black holes as Dark Matter as proposed in \cite{Calza:2024fzo} within the above theory framework.

\section*{ACKNOWLEDGEMENTS}

This work was partially supported by the program Unidad de Excelencia Maria de Maeztu CEX2020-001058-M, Spain (S.D.O).

\end{document}